\documentclass[12pt]{article}
\begin{document}

%%%%%% %%%%%%%%%

%%%%%%%%  %%%%%%%%  %%%%%%%%%

\begin{center}
{\LARGE{ Large $N$ field theories from Superstrings \footnote{
Lectures at the Latin American School SILAFAE III (April 2000) Cartagena, Colombia.} }}
\vskip 14mm

{\bf {\large{Jorge G. Russo }}}

\vspace{12 mm}

{\it Departamento de F\'\i sica, Universidad de Buenos Aires\\
Ciudad Universitaria, 1428 Buenos Aires}\\
{\tt russo@df.uba.ar }\\

\end{center}

\vskip 8mm

\begin{center}

{\bf Abstract}

\end{center}

These are   introductory lectures on the correspondence between $SU(N)$ gauge theories and Superstring Theory in anti-de Sitter
geometries (AdS).
The subject combines a number of different topics, including supersymmetric field theory,
classical and quantum physics of black holes, string theory, string dualities,
conformal field theories (CFT), and quantum field theory in anti-de Sitter spaces.
We also discuss applications of this AdS/CFT correspondence to  the large $N$ dynamics of  pure QCD.

\vspace{2 mm}

\newpage

\def\be{\begin{equation}}
\def\ee{\end{equation}}
\def\ba{\begin{array}}
\def\ea{\end{array}}

\def\a{\alpha }
\def\ads{$AdS_5\times S^5$}
\def\p{\partial }

\def\tr{ {\rm Tr\ } }

%%%%%%%%%%%%%%%%%%%%%%%%
 \section{Introduction}
%%%%%%%%%%%%%%%%%%%%%%%%%

The  idea that  non-abelian gauge theories can be described in terms
of a string theory  \cite{schw,hoof,poly}
was originally motivated by the duality symmetry
of scattering amplitudes and by the appearance of a great number of hadronic resonances
in strongly coupled QCD, which
persisted even for  high values of the spin.
The mass squared of the lightest particle of a given spin $J$ was observed
to be proportional to $J$, i.e. $M^2=const. J$. This  simple law successfully
predicted the masses of  a few  resonances, and it soon became clear that
such spectrum of particles follows from the simple assumption that mass and
angular momentum come from a spinning relativistic string.

It took many years to find a precise implementation of this idea.
In 1997 Maldacena conjectured a duality between gauge theories and superstring theories propagating on certain backgrounds \cite{malda},
which was later formulated in more detail in \cite{GKP,witten}.\footnote{
An extensive review on the AdS/CFT correspondence, which includes a more complete reference
list, can be found in \cite{review}. Other recent reviews include \cite{dive,duff}.}
The simplest example of this duality involves supersymmetric $SU(N)$ Yang-Mills theory
in 3+1 dimensions with four supercharges (${\cal N}=4$)
and superstring theory compactified in a space $AdS_5\times S^5$, where
$AdS_5$ is five dimensional anti de Sitter spacetime, and $S^5$ represents a 
five sphere. Other examples involve superstring theory compactified in  singular ten dimensional
spacetimes and Yang Mills theories in $p+1$ dimensions \cite{itza}.
It is now widely expected that a similar correspondence should hold for any
quantum gravity theory in a spacetime that is asymptotic to anti-de Sitter space.

Such correspondence also represents a concrete implementation \cite{witten} of another remarkable
conjecture by 't Hooft \cite{thoof}.
Based on the entropy formula of black holes, 't Hooft conjectured 
that a consistent quantum theory of gravity must be holographic, in the sense that
the fundamental degrees of freedom describing physics in a given volume can be ascribed
to the surface enclosing that volume. Moreover, there must be one degree of freedom
for each Planck unit area. The holographic idea was subsequently discussed
by Susskind in \cite{susskind} and the basic motivation comes from the Bekenstein bound \cite{beke},
which asserts that the black hole entropy proportional to the area  in Planck units is
the maximum physically possible entropy for any system that can be placed in
the same region.

Here we give a review on the subject and point out some attempts to exploit these ideas to understand ordinary QCD.
Instead of presenting a systematic and detailed  description of the many topics which are involved in this subject --which is clearly impossible  in two lectures-- we will try to provide
just the  basic ingredients of each field that are needed in order to understand the way the
AdS/CFT correspondence works, and how to use it to make predictions for strongly coupled gauge theories.

These lectures are  organized as follows. We will first briefly discuss the $1/N$ expansion in gauge theories \cite{cole}, which is one of the most compelling arguments that support a connection with string theory (section 2).
We then introduce the different string theories and string dualities (section 3), and discuss
brane solutions in string theories and the role of anti-de Sitter spaces (section 4).
Particular emphasis is made on the properties of D branes in string theory.
Some elements of conformal field theories and ${\cal N}=4$ supersymmetric Yang-Mills theory
are given in section 5. In section 6 and 7 we describe the
AdS/CFT correspondence 
and holography. In section 8 we describe some attempts to use the correspondence
to compute glueball spectrum and other properties of strongly interacting non-supersymmetric QCD.

%%%%%%%%%%%%%%%%%%%%%%%%%%
\section{$1/N$ expansion in gauge theories}
%%%%%%%%%%%%%%%%%%%%%%%%%

Despite numerous progresses in the description of QCD since it was proposed as a theory
of strong interactions, we do not yet have an analytic control or a detailed understanding
of basic phenomena, such as quark confinement.

The basic idea of the $1/N$ expansion, introduced by 't Hooft \cite{hoof}, is to consider the non-abelian gauge theory with gauge group $SU(N)$, and express physical quantities as a systematic expansion  in powers of $1/N$.
The  interactions between hadrons is expected to be a $O(1/N)$ effect.
This means that  in the $N=\infty $ limit  one can consider the problem of confinement
and hadron mass spectrum without the complication of residual hadronic interactions.

At large $N$, QCD is expected to be a string theory. This is supported by
the following facts:

\noindent (1) Feynman diagrams are organized in a topology expansion, just like string worldsheets.

\noindent (2) Experiments indicate a string-like behavior  (resonances and duality symmetry
of scattering amplitudes  between different channels).

\noindent In addition, 
 QCD contains string-like objects, which are the electric flux tubes between quarks and
antiquarks. The energy increases linearly
with the quark-antiquark distance, thus causing confinement.

Maldacena conjecture prescribes which is exactly the string theory corresponding to $SU(N)$
supersymmetric Yang-Mills theory with four supersymmetry generators:
it is type IIB superstring theory on the space $AdS_5\times S^5$ with string coupling $g=const. 1/N$.

The mechanism by which the $1/N$ expansion leads to an expansion in topologies of a two-dimensional space applies not only to gauge theories but to more  general
models.
To illustrate this point and see how the $1/N$ expansion works,  we consider a general field theory, with  degrees of freedom $A_m$ transforming in the adjoint of  $U(N)$,
$A_{ma}^{\ \ b}=- A_{mb}^{*\  a}$, which do not need to be scalar fields 
(the index $m$ could be a Lorentz index). The action is assumed to be of the form
$$
S=\int d^dx \bigg({\rm tr}\big[ \p A_m \p A_m \big]+g f_{mnp} {\rm tr}\big[ A_m A_n A_p\big]
$$
\be
+\ g^2 h_{mnpr} {\rm tr}\big[ A_m A_n A_p A_r\big]\bigg)\ .
\label{uno}
\ee
Here $f_{mnp}$ and $h_{mnpr}$ are arbitrary couplings, which do not depend on $g$ and $N$.
The interaction terms may also involve derivatives of the field, for example,
a term $g f_{mnp} '{\rm tr}\big[ A_m A_n \p A_p\big]$.
The only thing that will matter for the $N$ dependence of Feynman diagrams
 will be the way the coupling constant $g$ appears in (\ref{uno}), and the fact that  
$A_{ma}^b$ are in the adjoint.
Let us introduce 
$$
\lambda=g^2 N\ .
$$
We now consider the limit $N\to \infty $ and $g^2\to 0$ with fixed $\lambda $.
It is convenient to rescale $A_m$ as $A\to A/g$, so that the action takes the
form
$$
S={N\over\lambda }\int d^d x \bigg( {\rm tr}\big[ \p A_m \p A_m \big]+
f_{mnp} {\rm tr}\big[ A_m A_n A_p\big]
$$
\be
+\ h_{mnpr} {\rm tr}\big[ A_m A_n A_p A_r\big]\bigg)\ .
\label{dos}
\ee
In this form, we see that a general Feynman diagram will have
a factor ${N\over\lambda }$ for each vertex and a factor ${\lambda\over N}$ for each propagator.
In addition, for each loop of group indices, there will be a factor $N$, coming from
$\sum_a \delta^a_a=N$. Thus, if we denote by $V,E,F$ the number of vertices, propagators
and group index loops, respectively, each Feynman diagram will carry a factor
$$
K= N^{V-E+F}\ .
$$
This $N$ dependence dictates what are the dominant Feynman diagrams.
It is convenient to represent the propagator by a double line with opposite arrows \cite{hoof}
(i.e. viewing the adjoint representation as a direct product of fundamental and antifundamental representations). In this form, each Feynman diagram can be viewed as a simplex,
where $F$ is the number of faces, $E$ is the number of edges, and $V$ is the number of
vertices. A theorem due to Euler relates the combination $V-E+F$ to the genus $H$ (i.e. the number of handles) of the
surface:
$$
V-E+F=2-2H\ .
$$
Thus the perturbative expansion will be organized as a sum over topologies of
a two-dimensional surface. For example,
a vacuum amplitude will be given by a sum of the form
$$
A(N,\lambda )=\sum_{H=0}^\infty N^{2-2H} A_H(\lambda )\ .
$$
In the limit of large $N $, only the leading term survives. This corresponds to surfaces
with no handles. These are the planar diagrams.

The fact that  in the large $N$ limit the perturbative expansion of gauge theories
is naturally organized in an expansion in two dimensional topologies indicates
that at $N\gg 1$ gauge theories should admit a string theory description.
Presently, only a few concrete examples of this correspondence between string and gauge theories are known, as discussed below.

%%%%%%%%%%%%%%%%%%%%%%%%%%%%%%%%%%%%%%
\section{String theory}
%%%%%%%%%%%%%%%%%%%%%%%%%%%%

%%%%%%%%%%%%
\subsection{Generalities}
%%%%%%%%%%%%%%%%%

Let us recall the basic features of string theory \cite{Polchinski}.
Superstring theory unifies gravity and gauge forces in a consistent quantum theory.
Quantum states are described by excitations of strings.
%  Some string theories contain both open or closed strings,
% but some string theories contain only closed string excitations 
% (in the absence of D-branes, see below).
The theory has a single fundamental dimensionfull scale $\a' =l_s^2$  (with dimension length$^2$) and $T=1/(2\pi \a' )$ is the string tension.
The masses of string excitations are proportional to $1/l_s$.
At low energies, superstring theories reduce to a quantum field theory of spin $\leq 2$
particles (i.e. Einstein theory coupled to other particles).
They have the property of supersymmetry. We remind that this is a symmetry of the theory
involving transformations in spacetime, where the supersymmetry generators are  spinors, and
 two supersymmetry transformations amount to
a spacetime translation.

One can classify supersymmetric field theories according to the number of supersymmetry generators.
The maximum number of supersymmetry generators that a relativistic theory can have is $32$.
The basic reason for this bound is that theories with a higher number of supersymmetries necessarily
involve a massless particle with spin greater than 2 in four dimensions. 
It is believed that a relativistic theory for such particles does not exist.
This also implies that supersymmetric field theories can be defined only in $d\leq 11$,
since for higher dimensions spinors have more than 32 components.

In $d=11$, there is only one supersymmetric field theory, where the supersymmetry generator
is a Majorana spinor of 32 components. This theory contains a spin 2 particle (the graviton)
and it is known as eleven-dimensional supergravity.
The theory is not renormalizable as a quantum field theory by perturbative expansion
around flat space. It is believed to describe the low-energy regime of a consistent quantum theory
called M-theory.

In $d=10$, there are two theories with $32$ supersymmetries, type IIA and type IIB supergravity.
The supersymmetry generators are two Majorana-Weyl spinors of 16 components; in the case of type IIA supergravity they have the opposite chirality, while for type IIB they have the same chirality.
They describe the low energy regime of {\it type IIA } and {\it type IIB superstring theories}.

In $d=10$ one can also have  a theory with 16 supersymmetries, i.e. a single Majorana-Weyl
supersymmetry generator. There are two types of super multiplets, the gravity multiplet (leading to type I supergravity) and the Yang-Mills
multiplet (leading to 10d super Yang-Mills theory). The theory is anomaly free only if
both theories are coupled together, and the gauge group is $E_8\times E_8$ or $SO(32)$.
In the first case, the theory describes the low-energy regime of
$E_8\times E_8$ {\it heterotic string theory}.
In the second case, the theory describes  the low-energy regime of either {\it type I superstring theory} or $SO(32)$ {\it heterotic string theory}.
These five superstring theories are related by duality symmetries (see below).

Let us consider the type IIA and type IIB superstring theories.
The 32 supercharges transform under the Lorentz group as the direct sum
of two Majorana-Weyl spinor representations, 
$$
{\bf 32}={\bf 16}+ {\bf 16}'  \ .
$$
The massless bosonic degrees of freedom are separated in two sectors called
NSNS and RR sectors. In both type IIA and IIB cases the NSNS sector
includes a metric $g_{\mu\nu}$, an antisymmetric two-form gauge potential
$B_{\mu\nu}$, and a scalar (dilaton) field $\phi $.
The RR sector of type IIA theory contains  a one-form and a three-form gauge potentials
$\{ A_\mu, A_{\mu\nu\rho } \} $.
The RR sector of type IIB theory consists of  a  pseudo-scalar field $A$, a two-form 
$A_{\mu\nu}$ and a four-form field $A_{\mu\nu\rho\sigma  }  $ with self-dual field strength.

The low energy dynamics of the massless fields is governed by the effective action
$$
S={1\over 16\pi G_{10} } \int d^{10}x \sqrt{-g} e^{-2\phi } \bigg[ R+4 (\p \phi )^2 
$$
\be 
-{1\over 12}
H_{\mu\nu\rho }H^{\mu\nu\rho}\bigg]+ ( {\rm RR\ fields})
\ee
\be
16\pi G_{10}=(2\pi )^7 g^2 l_s^8\ ,\ 
\label{nombre}
\ee
$$
H_{\mu\nu\rho}=\p_\mu B_{\nu\rho }+\p_\nu B_{\rho\mu }+
\p_\rho B_{\mu\nu }\ ,
$$
where $g$ is the string coupling constant. Note that the string coupling constant
is determined by the expectation value of the dilaton field, 
$$
g=e^{\phi_0}  \ .
$$
There are also higher derivative terms which can be neglected at low energies.

%%%%%%%%%%%%%%%%%%%%%%%%%%%
\subsection{String dualities}
%%%%%%%%%%%%%%%%%%%%%%%%%%%%%%

A duality is an equivalence between theories that are seemingly different.
In particular, the theories may have  different fields, but nevertheless the same spectrum and  amplitudes.
Typically, duality symmetries appear when a quantum system has two different
classical limits.

An important duality symmetry of string theory is T-duality. It has no analogue in field theory.
Consider for example that the dimension $x_{10}$ is compactified on a circle $S^1$,
$$
x_{10}=x_{10}+2\pi R\ .
$$
String states will be specified by a set of  quantum numbers, but those relevant to
T-duality are
$w,p$, where $w$ is the number of times the string winds around $S^1$,
and $p$ represents the quantized (integer) momentum in the $x_{10}$ direction.
T-duality is the remarkable property that physics is invariant under the
simultaneous exchange of $R\leftrightarrow \a'/R$  and $p\leftrightarrow w$.
Thus this symmetry relates a compactification on a small distance with a compactification
on a large distance.

A T-duality transformation changes the chirality of left handed fermions (or right handed fermions, depending on the conventions). As a result, it connects type IIA superstring theory
to type IIB superstring theory.
Thus type IIA superstring theory compactified on a circle of radius $R$ is equivalent
to type IIB superstring theory compactified on a circle of radius $\a' /R$.
Winding states in one description correspond to momentum states in the other.

Another important duality symmetry is $S$-duality, which relates weak $g\ll 1$ and strong 
$g\gg 1$ coupling regimes. The strong coupling limit of string theory is in general complicated, but one can infer a
possible S-duality symmetry from the effective field theory.
A crucial check is the spectrum of supersymmetric states, which can be compared
to the dual candidate, since it does not receive quantum corrections, i.e. it can be extrapolated
to $g\gg 1$.
The basic reason can be understood schematically as follows.
Let $Q$ be a supersymmetry generator, and $Z$ a  gauge symmetry generator, so that its eigenvalues represent some charge. Because they are symmetry generators, they must commute with the hamiltonian $H$,
$$
[H,Z]=0\ ,\ \ \ \ [Q,H]=0\ .
$$
The supersymmetry algebra is of the form
$$
Q^2=H-Z\ .
$$
Consider a quantum state $| \Psi \rangle $ of energy $E$ and charge $q$, which is supersymmetric, i.e. which is annihilated by $Q$,
 $Q| \Psi \rangle =0$. Thus
$$
0=\langle \Psi | Q^2 |\Psi \rangle =\langle \Psi | H |\Psi \rangle -\langle \Psi | Z |\Psi \rangle =E-q
$$
Hence $E=q$. This equality (called the BPS condition) holds independently of the value
of $g$, since the above derivation relied only on symmetries. Therefore it must hold for $g\gg 1$ as well.

By introducing a new (``Einstein frame") metric $g^E_{\mu\nu }=e^{-\phi/2} g_{\mu\nu }$,
the low-energy action of type IIB superstring theory takes the form
$$
S_{\rm IIB}=\int d^{10}x \sqrt{-g_E}\bigg[ R(g_E)-{1\over 2} (\p\phi)^2 
$$
$$
-{1\over 12} e^{-\phi } H_{\mu\nu\rho} H^{\mu\nu\rho}
-{1\over 12} e^{\phi }\tilde H_{\mu\nu\rho} \tilde H^{\mu\nu\rho}\bigg]+...
$$
$$
H_3=dB_2\ ,\ \ \ \ \ \tilde H_3=dA_2
$$
This action is manifiestly symmetric under
$$
g^E_{\mu\nu}\to g^E_{\mu\nu}\ ,\ \ \ \ \ \phi\to -\phi\ ,\ \ \ \ \ 
H_3\leftrightarrow \tilde H_3\ .
$$
This transformation exchanges $g=e^{\phi _0}$ by $1/g=e^{-\phi_0}$, and thus
it exchanges weak and strong coupling regimes.
This indicates that type IIB superstring theory with coupling $g$ is equivalent to type IIB
superstring theory with coupling $1/g$, a conjecture that  passed numerous non-trivial
 tests.
Note that this duality exchanges states with NSNS and RR charges.

The full S-duality symmetry group of ten-dimensional type IIB superstring theory
is the $SL(2,Z)$ group, which includes, apart from the above transformation, shifts in the
RR scalar field $A$. More precisely, defining
$$
\tau ={A\over 2\pi } +i e^{-\phi }=\tau_1+ i\tau_2
$$
an $SL(2,Z)$ transformation acting on $\tau $ is of the form
$$
\tau \to {a\tau +b\over c\tau +d}\ ,\ \ \ \ ad-bc=1\ ,\ \ a,b,c,d \ \in  Z
$$
This is reminiscent of a similar $SL(2,Z)$ symmetry of ${\cal N}=4$ super Yang-Mills theory,
with parameter $\tau ={\theta\over 2\pi } +i{4\pi \over g_{\rm YM}^2}$. As we shall see below, this symmetry of super Yang-Mills theory  can be explained from the $SL(2,Z)$ symmetry of type IIB theory by
the Maldacena conjecture.

The $SL(2,Z)$ symmetry has also a simple geometrical interpretation. Type IIB theory 
arises as dimensional reduction of the eleven dimensional M-theory
compactified on a 2-torus in the limit the torus area goes to zero with fixed modular parameter.
Then the $SL(2,Z)$ duality symmetry of type IIB superstring theory is identified with the
modular symmetry of the 2-torus.

%%%%%%%%%%%%%%%%%%%%%%%%%%%%%
\section{Black holes and black $p$ branes}
%%%%%%%%%%%%%%%%%%%%%%%%%%%%%%%

%%%%%%%%%%%%
\subsection{Reissner-Nordstrom black hole solution in $d=4$}
%%%%%%%%%%%%%%%

In four-dimensional Einstein-Maxwell theory, a general stationary solution is fully characterized
by mass $M$, charge $q$ and angular mometum $J$.
For $J=0$, one has the Reissner-Nordstrom black hole with metric ($G=1$)
\be
ds^2=-\lambda (r) dt^2+ {1\over \lambda(r)} \ dr^2 +r^2 d\Omega_2^2\ ,\ 
\ee
$$
d\Omega_2^2=d\theta^2+\sin^2\theta d\varphi ^2 \ ,
$$
$$
\lambda(r)=1-{2M\over r}+{q^2\over r^2}={1\over r^2} (r-r_+) (r-r_-)
$$
$$
r_\pm =M\pm \sqrt{M^2-q^2}\ ,\ \ \ \ M\geq q\ ,\ \  
$$
$$
q=\int_{S^2}\epsilon _{\mu\nu}^{\ \ \ \rho\sigma}F_{\rho\sigma} \ dx^\mu dx^\nu 
$$
If $M=q$, then the inner and outer horizon coincide, $r_+=r_-$, and we have
an extremal black hole. Let us examine the near-horizon geometry of the extremal black hole. The metric takes the form
\be
ds^2 \cong -e^{2\rho/r_+} dt^2 +d\rho^2 +r_+^2 d\Omega_2^2
\label{wads}
\ee
$$
\rho=r_+ \log \big( {r\over r_+} -1 \big)
$$
This describes the direct product of a two dimensional anti-de Sitter space-time $AdS_2$
and a 2-sphere, i.e. $AdS_2\times S^2$.

%%%%%%%%%%%%%%%%%%%%%%%%%%%%%%%
\subsection{Anti-de Sitter space}
%%%%%%%%%%%%%%%%%%%%%%%%%%%%%%%

The anti-de Sitter space is a maximally symmetric spacetime with constant negative curvature.
The space $AdS_2$ that emerged in (\ref{wads}) can be represented in ${\bf R}^3$ by a hyperboloid
$$
x_0^2+x_2^2-x_1^2=r_+^2
$$
More generally, 
the $(p+2)$-dimensional anti-de Sitter space $AdS_{p+2}$ is the hyperboloid
\be
x_0^2+x_{p+2}^2-\sum_{i=1}^{p+1} x_i^2= R^2
\label{hype}
\ee
where $x_0, x_{p+2}, x_i $ are coordinates of ${\bf R}^{p+3}$ with metric
\be
ds^2=-dx_0^2-dx_{p+2}^2+\sum_{i=1}^{p+1} dx_i^2
\label{qqq}
\ee
In this representation, it is clear that the isometry group is $SO(2,p+1)$.
The space $AdS$ is homogeneous and isotropic and it has the maximal number of Killing vectors (equal to ${1\over 2}(p+2)(p+3)$).

The {\it Poincar\' e metric} of $AdS$ can be obtained by writing
$$
R^2 z^{-1}=x_{p+1}+x_{p+2}\ ,\ \ \ \ v=-x_{p+1}+x_{p+2}\ ,\ 
$$
$$
 z_n={z\ x_n\over R}\ .
$$ 
Inserting into (\ref{qqq})\ we get
\be
ds^2={R^2\over z^2}\big[ dz^2-dz_0^2+dz_1^2+...+dz_p^2\big]\ .
\ee
Because the metric is conformal to a flat metric, the Weyl tensor of this space is identically zero. For the Riemann tensor one has
\be
R_{\mu\nu\rho\sigma }={1\over R^2} (g_{\mu\rho}g_{\nu\sigma }
-g_{\mu\sigma} g_{\nu\rho})\ .
\ee
Hence
$$
R_{\mu\nu }=-{p\over R^2} \ g_{\mu\nu}\ .
$$
By introducing a coordinate $u=1/z$, the $AdS$ metric can also be written as
\be
ds^2={R^2u^2}\big[ -dz_0^2+dz_1^2+...+dz_p^2\big] +R^2{du^2\over u^2}\ .
\label{ades}
\ee

The space $AdS_{p+1}$ has closed timelike curves, which can be removed by
considering the universal cover (see e.g. \cite{hawell}).
Then the boundary of $AdS_{p+1}$ is topologically $S^p\times {\bf R}$.
This can be seen
from the hyperboloid representation (\ref{hype}). In terms of the coordinates of metric (\ref{ades}), the boundary
is constituted of two components, a single point  at $u=0$, and $u=\infty $, which is the Minkowski space ${\bf R}^{p+1}$. Thus the boundary of $AdS_{p+1}$ is a conformal
compactification of the Minkowski space ${\bf R}^{p+1}$.

%%%%%%%%%%%%%%%%%%%%%%%%%%%%%%
\subsection{Black p-branes and D-branes}
%%%%%%%%%%%%%%%%%%%%%%%%%%%%%%

p-branes are extended objects which can be classified by their charges.
A 0-brane represents a point-like particle, a 1-brane represents a string-like object,
a 2-brane represents a membrane, etc.
Consider standard electromagnetism in four dimensions. For point-like configurations
one can define electric and magnetic charges as follows:
$$
Q_{\rm elec}=\int_{S^2} \ *\ F_2=\int _{S^2} \epsilon _{\mu\nu}^{\ \ \ \rho\sigma}F_{\rho\sigma} \ dx^\mu dx^\nu \ ,
$$
$$
Q_{\rm mag}=\int_{S^2} \  F_2=\int _{S^2} \ F_{\mu\nu } \ dx^\mu dx^\nu \ .
$$
Just as a point particle or 0-brane couples to a one-form gauge potential 
$A_1=A_\mu dx^\mu $, a p-brane couples to a $p+1$ gauge potential
$A_{p+1}$, with field strength $F_{p+2}=dA_{p+1}$, and electric and magnetic charges are defined by ($d=10$)
$$
Q_{\rm elec}^{(p)}=\int_{S^{8-p}} \ *\ F_{p+2}\ ,
$$
$$
Q_{\rm mag}^{(p)}=\int_{S^{p+2}} \  F_{p+2}\ .
$$
In ten dimensions, the electrically charged object describes a $p$-brane, and the magnetically charged object describes a $6-p$ brane.

We have seen that in type IIA and type IIB theory there is an antisymmetric two-form
 $B_{\mu\nu}$ in the NSNS sector (which will give rise to black strings and
 magnetic dual black five branes) and there are in addition various gauge fields
in the RR sector. Objects which carry RR charges are called D-branes.
The corresponding geometries are obtained by solving the equations of motion
of the low-energy string-theory effective action
$$
S={1\over 16\pi G_{10} }\int d^{10}x \sqrt{-g} \bigg( e^{-2\phi }\big[ R+ 4 (\p \phi )^2\big]
$$
\be
-\ {2\over (8-p)!} F^2_{p+2} \bigg)\ .
\ee
The extremal Dp-brane background is given by
$$
ds^2= f^{-1/2}(r)\big[ -dt^2+dx_1^2+...+dx_p^2\big]
$$
\be
+\ f^{1/2}(r) (dr^2+r^2 d\Omega_{8-p}^2\big)\ ,
\label{kkk}
\ee
\be
e^{2\phi}=g^2 f^{ {3-p\over 2}}\ ,\ \ \ \ f(r)=1+{R^{7-p}\over r^{7-p}}\ ,
\ee
$$
\int _{S^{8-p}} * F_{p+2}=N\ ,\ \ \ \ R^{7-p}=c_p g N l_s^{7-p}\ ,\ 
$$
$$
c_p=2^{5-p}\pi^{5-p\over 2} \Gamma ({7-p\over 2})\ .
$$
The extremal Dp brane has  mass 
\be
{ {\rm mass}\over V_p}={N\over (2\pi )^p l_s^{p+1}}\ {1\over g}\ ,
\ee
where $V_p$ represents the volume of the Dp-brane.
What makes a D-brane special is the fact that in string units the mass is proportional to $1/g$.
Since the gravitational field produced by an object is proportional to the mass times
the Newton constant $G_{10}\sim g^2$, this means that it vanishes as $g\to 0$.
This indicates that for $g\ll 1$ D-branes must admit a flat theory description.
Such description was found by Polchinski \cite{pol}, and it has led to many important results.
The observation of \cite{pol} is that, 
in the limit $g\ll 1$, a D-brane can be represented by a $(p+1)$-dimensional hyperplane
defined as a place where open strings can end (Dirichlet branes).
If there are several hyperplanes, there can be open strings with ends attached to different
D-branes.
It can then be shown that $N$ Dp-branes carry exactly $N$ units of $(p+1)$-form
RR charge.

The low energy effective theory of open strings on 
$N$ coinciding Dp-branes (obtained by taking the limit $\a'\to 0$) is $U(N)$ gauge
theory in $p+1$ dimensions with 16 supersymmetry generators \cite{witten95}.
The gauge coupling is related to the string coupling by
$g_{\rm YM}^2=4\pi g$.
As $g$ is increased, the gravitational field of Dirichlet branes grows and they eventually become black $p$ branes.
The connection between low energy D-branes and gauge theories indicates that,
in a suitable low-energy limit, it should be possible to describe black branes by a strongly coupled field theory.

If one D brane is separated from the others the gauge group is broken as
$U(N)\to U(N-1)\times U(1)$.
There are $2(N-1)$ gauge bosons which get a mass. These are the strings
which go from the $N-1$ D-branes to the D-brane that was separated
(the factor 2 arises because there are two possible orientations).

%%%%%%%%%%%%%%%%%%%%%%%%%
\section{ Conformal field theories and ${\cal N}=4$ SYM}
%%%%%%%%%%%%%%%%%%%%%%%%

%%%%%%%%%%%%%%%%%%
\subsection{Conformal group}
%%%%%%%%%%%%%%%%%%%

The conformal group is constituted of transformations that preserve
the metric up to a scale factor, $g_{\mu\nu}(x)\to g_{\mu\nu} \Omega^2(x)$.
This group incorporates Poincar\' e transformations and scale transformations.
The generators are the usual Lorentz generators $M_{\mu\nu}$, the Poincar\' e
translation operators $P_\mu $, and in addition generators $D$ and $K_\mu $.
The conformal group is isomorphic to $SO(d,2)$, with the identification
$$
{\cal M}_{\mu\nu}=M_{\mu\nu}\ ,\ \ \ \ \ {\cal M}_{ d\mu}={\textstyle {1\over 2}}
(P_\mu-K_\mu )\ , \ \ 
$$
$$
{\cal M}_{\mu (d+1)}={\textstyle {1\over 2}}(P_\mu +K_\mu )\ ,\ \ \ \ {\cal M}_{d(d+1)}=D \ .
$$
The scaling dimension $\Delta $ of an operator $\varphi (x)$ is dictated by the transformation rule under scaling of coordinates:
$$
D:\ \ \ \ x^\mu \to \lambda x^\mu\ ,\\ \ \ \ \varphi(x)\to \varphi'(x)=\lambda ^\Delta \varphi (\lambda x)
$$
{\it Primary operators} are the lowest dimension operators and they are annihilated
by $K_\mu $ at $x^\mu =0$.
Representations of the conformal group are labelled by the scaling dimension
$\Delta $ and the Lorentz representation.

Two and three-point correlation functions of primary fields are entirely  determined by conformal symmetry. For example
$$
\langle \varphi(x) \varphi (x')\rangle =const. {1\over |x-x'|^{2 \Delta} }
$$
To combine the conformal algebra with the supersymmetry algebra, 
additional fermionic generators $\tilde Q $ must be included, which arise from $[K,Q]\sim \tilde Q$.
As a result, the number of fermionic generators in the superconformal algebra is doubled with respect to the non-conformal case.
For example, for a field theory with particles of spin $\leq 1$,
the maximal number of supercharges of the supersymmetry algebra is 16,
and the maximal number of fermionic generators in a superconformal field theory is 32.

%%%%%%%%%%%%%%%%
\subsection{ ${\cal N}=4$ super Yang-Mills theory}
%%%%%%%%%%%%%%%%%%%%%%%%%%%

Non-supersymmetric 3+1 dimensional pure Yang-Mills theory is scale invariant, but it has $\beta \neq 0$
at quantum level. 
An example of superconformal field theory is ${\cal N}=4$ $U(N)$ super Yang-Mills theory
in 3+1 dimensions, which has exact conformal invariance
($\beta =0$ to all orders).
 It contains 16 supercharges, which under the Lorentz group transform as four spinors
$(Q_\a^A,\bar Q_{\dot \beta} ^A)$,\ $A=1,2,3,4$, where $Q_\a $, $\bar Q_{\dot\beta }$ are Weyl spinors.
An  $SU(4)$ rotation of the four spinors is an automorphism of the supersymmetric algebra.
As a result, the Lagrangian is invariant under $SU(4)\sim SO(6)$ global transformations
(R-symmetry).

The degrees of freedom of the theory are as follows:

\smallskip
\noindent i)
A vector field $A_\mu $ in the adjoint representation of $SU(N)$ which is a singlet under $SO(6)$.

\smallskip
\noindent ii)
Six real scalars $X^a$ in the ${\bf 6}$ vector representation of $SO(6)$, which
transform in the adjoint representation of $SU(N)$.

\smallskip
\noindent iii)
Four Weyl fermions $\lambda_\a^A$ transforming in the adjoint of $SU(N)$ and ${\bf 4}$ spinor representation of $SO(6)$ (corresponding to the fundamental representation  of $SU(4)$).

\smallskip
\noindent 
The lagrangian of the theory can be derived by dimensional reduction of $d=10$ ${\cal N}=1$
super Yang-Mills theory:
\be
{\cal L}=-{1\over 4 g_{\rm YM}^2} \tr \big[ F_{MN} F^{MN} \big]-{i\over 2}\tr
\big[ \bar\lambda \Gamma^M D_M \lambda \big]
\ee
Here $\lambda $ is a Majorana-Weyl ${\bf 16}$ spinor of $SO(1,9)$.
Upon reduction, we have the decomposition 
$$
SO(1,9)\to SO(1,3)\times SO(6)
$$
under which
$$
{\bf 16} =(2,4)+(\bar 2,\bar 4)
$$
The ten-dimensional gauge field gives rise to a 4d gauge field plus six scalar fields:
$$
A_M=(A_\mu, X_a)\ ,\ \ \ \ 
M=(\mu, a)\ ,
$$
$$
 \mu=0,1,2,3\ , \ \ \ \ \ a=4,...,9\ .
$$
The dimensionally reduced Lagrangian is then obtained as usual by  assuming that 
fields depend only on $x^\mu$. It takes the form
$$
{\cal L}=-{1\over 4 g_{\rm YM}^2} \tr \bigg( F_{\mu\nu } F^{\mu\nu } +2D_\mu X_a D^\mu X_a 
$$
\be
-[X_a,X_b]^2\bigg)
-\ {i\over 2}\tr
\bigg(\bar\lambda \Gamma^\mu D_\mu \lambda +i\bar \lambda \Gamma_a [X_a,\lambda ] \bigg)
\ee

%%%%%%%%%%%%%%%%%%%%%%%%%%%%%%%%%%%%%%
\section{AdS/CFT correspondence}
%%%%%%%%%%%%%%%%%%%%%%%%%%%%%%%%%%%%%%

We now have the basic elements to understand the correspondence
between string theory on $AdS$ spaces and conformal field theories.
Let us consider type IIB superstring theory in the presence of $N$ D3 branes.
String theory on this background contains two types of excitations, closed
and open strings. Closed strings propagate on the bulk, whereas open strings
are attached to the D3 branes and go from one D3 brane to another.

For energies much less than $1/l_s$, only massless excitations appear.
The effective action for massless fields is of the form
$$
I=\int d^{10}x \ {\cal L}_{\rm IIB} + \int d^4x\ {\cal L}_{\rm brane}\ ,
$$
where ${\cal L}_{\rm IIB}$ is the effective lagrangian of type IIB string theory, which contains the supergravity action discussed in section 4 plus higher derivative terms,  and
${\cal L}_{\rm brane}$ is the lagrangian for the low-energy theory on the brane. 

Let us now take the limit $\a' =l_s^2\to 0$ (low energy limit). The gravitational coupling
is (see eq.~(\ref{nombre})~)
$$
8\pi G_{10}=\kappa ^2\sim g^2 {\a'}^4 \to 0\ .
$$ 
 Thus in this limit gravitational interactions and higher-derivative terms in the string effective action vanish. As discussed before, in the limit 
$\a' \to 0$, ${\cal L}_{\rm brane}$ reduces to the lagrangian of 
 $U(N)$ super Yang-Mills theory in $d=3+1$ dimensions.
Thus the resulting theory is ${\cal N}=4$ $U(N)$ SYM and free gravity propagating on the bulk
spacetime.

Let us now make use of the supergravity description for the same configuration. 
We have seen in section 4 that
D-branes are extended objects with RR charges which, having a nonzero mass, produce a gravitational field.
They are described by the solution (see eq.~(\ref{kkk})
$$
ds^2= f^{-1/2}(r)\big[ -dt^2+dx_1^2+dx_2^2+dx_3^2\big]
$$
\be
+\ f^{1/2}(r) (dr^2+r^2 d\Omega_{5}^2\big)
\label{pppp}
\ee
\be
F_{0123r}=\p _r f^{-1}\ ,\ \ \ \ e^{2\phi}=g^2 ={\rm const}\ ,\ 
\ee
$$
 f(r)=1+{{\a' }^2R^4\over r^4 }\ ,
\ \ \ \ 
R^4=4\pi g N\ .
$$
There are two types of low-energy excitations in this background:
massless particles in the bulk, and excitations close to the horizon at $r=0$.
Indeed, because of the large redshift in the vicinity of the horizon, an excitation
near the horizon of energy $E$ is measured by an observer at infinity
with energy
\be
E_\infty =f^{-1/4} E  \sim {r\over \a'} \ (E\sqrt{\a' })
\label{qqpp}
\ee
Taking the same $\a'\to 0$ limit as in the previous description, with $r/\a' \equiv u $ fixed
(so $r\to 0$),  then a given string excitation of ``level" $n=a' E^2$  has a finite energy $E_\infty $ at infinity.
In this limit the metric (\ref{pppp}) becomes
$$
ds^2=\a ' \bigg[ {u^2\over R^2} \big(-dt^2+dx_1^2+dx_2^2+dx_3^2\big)
$$
\be
+\ 
R^2{du^2\over u^2}+ R^2 d\Omega_5^2\bigg]
\label{uuu}
\ee
Comparing with metric (\ref{ades}) we see that this is the space $AdS_5\times S^5$.
It represents the geometry of the extremal black D3 brane metric near the horizon (this is
similar to the result of section 4 where $AdS_2\times S^2$ arised as the near-horizon
geometry of the extremal Reissner-Nordstrom black hole).

In sum, there are two different descriptions of the same configuration:
\smallskip

\noindent a) the weak-coupling description of the D3 branes in terms of open strings with Dirichlet boundary conditions, which in the limit $\a' \to 0 $ led to ${\cal N}=4$ 
$U(N)$ SYM theory;

\smallskip
\noindent b) Superstring theory on a background D3 brane geometry, which in the same limit
led to
superstring theory on $AdS_5\times S^5$.

\smallskip
\noindent 
Therefore type IIB superstring theory on  $AdS_5\times S^5$ must be equivalent
to ${\cal N}=4$ $U(N)$ SYM theory in 3+1 dimensions.
This is the Maldacena conjecture \cite{malda}.
The parameter $N$ appears in the string theory through the radius of $AdS_5$ 
(which has the same radius as $S^5$), viz. $R^2=\sqrt{4\pi g N}$,
and the Yang-Mills coupling is given by $g_{\rm YM}^2=4\pi g$.

When the curvature is $\ll {1\over \a' }$, the string excitations can be ignored and
one simply has supergravity theory on $AdS_5\times S^5$. This is the case
for $R^2\gg 1$, which requires $gN\gg 1$.
Since $ gN={g^2_{\rm YM}N  \over 4\pi }$= 't Hooft coupling, the supergravity approximation
is valid at strong 't Hooft coupling.
Because $g<1$ (for perturbative string theory to apply), the condition $gN\gg 1$ 
implies $N\gg 1$.

An important test of the conjecture is that symmetries on each side match
exactly.
The type IIB superstring theory on $AdS_5\times S^5$ has the following symmetries:

\smallskip
\noindent 1) the $SL(2,Z)$ duality group described in section 3; 

\smallskip
\noindent 2) 32 supersymmetries, which are left unbroken by the $AdS_5\times S^5$ background;

\smallskip
\noindent 3) The $SO(2,4)$ isometry of $AdS_5$.

\smallskip
\noindent 4) The $SO(6)$ isometry of $S^5$.

\smallskip
\noindent This is precisely the symmetry of  ${\cal N}=4$ 
$U(N)$ SYM theory. Indeed,  it has duality symmetry $SL(2,Z)$ under transformations
of $\tau={\theta\over 2\pi } +i{4\pi \over g^2_{\rm YM} }$;
the $SO(6)$ R symmetry described in section 5.2; the $SO(2,4)$ conformal symmetry;
and 32 supersymmetry generators of the superconformal group.

Let us examine the validity of the various approximations  in detail.
It is convenient to choose units to set $\a' ={1\over \sqrt{g N}}$.
Then the gravitational coupling is 
$$
\sqrt{G_{10} }\cong g (\a')^2={1\over N}\ .
$$
Corrections due to massive string excitations (i.e. $\a' $ corrections) 
will be of order $O({1\over \sqrt{g N}})$. This is also seen from the fact that
the masses of the string states are $M^2_{\rm str}\sim \sqrt{g N}$
and go to infinity as $\lambda= g_{\rm YM}^2N\to \infty $.
The masses of Kaluza-Klein states of the sphere are of order $O(1/R_0)$, with 
$R_0^2=R^2 \a '=\sqrt{gN}\a' =1$. Thus $M_{KK}^2=O(1)$.
String loops will be of order $O(1/N^2)$.

The two descriptions, perturbative YM and supergravity, apply in different regimes:
classical gravity applies for $R^2\gg \a' $, and this requires $g^2_{\rm YM}N\gg 1$.
Perturbative Yang-Mills requires $g^2_{\rm YM}N\ll 1$.

%%%%%%%%%%%%%%%%%%%%%%%%%%%%%%%%%%%%%%%
\section{Field/Operator correspondence \\ 
and Correlation functions}
%%%%%%%%%%%%%%%%%%%%%%%%%%%%%%%%%%%%%%%%%

%%%%%%%%%%%%%%%%%
\subsection{CFT correlators from supergravity}
%%%%%%%%%%%%%%%%%%%%%%

Deformations of the super Yang-Mills lagrangian by adding
gauge invariant operators correspond to changing the asymptotic 
values of string fields at infinity.
 For example, consider the string coupling
$$
g=e^{\phi(\infty )}
$$
Changing $g={g_{\rm YM}^2\over 4\pi }$ amounts to add a marginal operator in the Yang-Mills theory of the form $\tr F^2$.

At the boundary $u=\infty $, the string fields are general functions of $x^\mu $, which
play the role of sources for operators in the Yang-Mills
field theory. Correlation functions are obtained by the prescription \cite{GKP,witten}
$$
\langle \exp \bigg[\int d^4 x \phi_0 (x) {\cal O}(x)\bigg] \rangle  = 
Z_{\rm string}\big(\phi_0(x) \big) 
$$
\be
\cong  \exp \big[ - I_{\rm sugra}(\phi_0 )\big]
\label{zzzz}
\ee
where
$$
\phi_0(x) =\phi(x,u)\bigg|_{u=\infty}
$$
So each field propagating in $AdS$ is in correspondence with a CFT operator.
For example, consider a massive scalar field in $AdS_{d+1}$ of mass $m$.
It can be shown \cite{witten} that this is associated with a CFT operator of scaling
dimension $\Delta $,
$$
\Delta={d\over 2} +\sqrt{ {d^2\over 4}+ R^2 m^2}\ .
$$

The relation (\ref{zzzz}) suggests a generalization of the AdS/CFT
correspondence to general string vacua
of the form $AdS_5\times X_5$. Equation (\ref{zzzz}) can indeed be viewed as
the definition of conformal field theory correlators
in terms of the string theory partition function on a general space of the form $AdS_5\times X_5$.
Most of these backgrounds do not preserve any supersymmetry, so they define non supersymmetric conformal field theories in four dimensions. More generally,
one can use any ten-dimensional string solution that looks near infinity like
$Y\times X$, where $Y$ is an Einstein manifold.
Finding the dual field theory in general cases is difficult, except for the cases which have a D-brane interpretation, whose low-energy theory is understood.

%%%%%%%%%%%%%%%%%%%%%%%%%%%%%%%%%%%%%%%%
\subsection{Black hole entropy and holography}
%%%%%%%%%%%%%%%%%%%%%%%%%%%%%%%%%%%%%%%%

Black holes obey the fundamental laws of thermodynamics with an entropy $S={{\rm Area}\over 4G}$.
In statistical mechanics, the entropy is derived from the logarithm of the number of states of energy $M$ (with given total charge and total angular momentum).
Classically, a black hole is completely characterized by mass, charge and angular momentum
and therefore it cannot have any entropy.
It is an old problem to understand what are the states of quantum gravity
--which are unobservable in the classical theory-- that provide the precise degrees of freedom
to derive  the black hole thermodynamical entropy by a statistical-mechanics counting.

The fact that for a black hole $S={{\rm Area}\over 4G}$ indicates that in quantum gravity
the number of states inside a given volume is proportional to the area of the surface enclosing such volume. This led 't Hooft to formulate its {\it Holographic principle}: nature should
be ``holographic" in the sense that there should exist a description in terms of degrees of
freedom living on the boundary of space.

The AdS/CFT duality is holographic, since gauge fields are degrees of freedom 
which ``live" on the boundary of $AdS_5$
(which is $\sim {\bf R}^4$), and they describe, just as a hologram, the interior.
This was worked out in detail in \cite{witten}, where this point of view was first advocated.
Holography not only requires that the bulk spacetime can be described
by degrees of freedom on the boundary, but it also prescribes that there
should be a single degree of freedom for each Planck unit area.
This point is also satisfied in the AdS/CFT correspondence  by virtue of the so-called UV/IR connection \cite{suswit}.

%%%%%%%%%%%%%%%%%%%%%%%%%%%%%%%%%%%%%%%
\section{Models for QCD}
%%%%%%%%%%%%%%%%%%%%%%%%%%%%%%%%%%%%%%%

%%%%%%%%%%%%%%%%%%%%%%%%%%%
\subsection{Non-supersymmetric gauge theories\\
from D branes}
%%%%%%%%%%%%%%%%%%%%%%%%%%%

The low-energy theory of a D4 brane is super Yang-Mills theory in $d=4+1$
dimensions. Suppose that the dimension $x_4$ is compactified on a circle $S^1$ of radius
$r_0$.
There are two possible boundary conditions for fermions:

\smallskip
\noindent
a) Periodic. This choice respects supersymmetry (since fermions and bosons obey the same
boundary conditions) and leads to ${\cal N}=4$ SYM theory in 3+1 dimensions.

\smallskip
\noindent 
b) Anti-periodic. This breaks supersymmetry completely. 
Fermions acquire masses of order $1/r_0$ and scalar particles get masses
by loop corrections.
The resulting theory is non supersymmetric $U(N)$ Yang-Mills theory in
$3+1$ dimensions with no matter fields.

Thus a supergravity background describing 4+1 Yang-Mills theory with antiperiodic fermions
can be used as a model of standard (non supersymmetric) large $N$ QCD.
This idea was exploited by Witten in \cite{witt}.
In order to construct the relevant supergravity background, we start with the non-extremal D4 brane metric with the euclidean time $\tau $ describing the $x_4$ dimension.
Because this is related to a finite temperature case (with temperature
$T_H=(2\pi r_0)^{-1}$), fermions obey
antiperiodic boundary conditions in the compact euclidean time dimensions,
$\tau=r_0 \theta ,\ \theta=\theta+2\pi $.
The metric and dilaton are given by
$$
ds^2={8\pi \lambda u\over 3 u_0} \bigg[ u^2 \big[ -dx_0^2+dx_1^2+dx_2^2+dx_3^2\big]
$$
\be
+{u^2\over 9 u_0^2} \big( 1-{u_0^6\over u^6} \big) d\theta ^2
+\  {du^2\over u^2(1-{u_0^6\over u^6})} + {1\over 4} d\Omega_4^2\bigg]\ ,
\label{metro}
\ee
$$
e^{2\phi }={8 \pi \lambda^3 u^3\over 27 u_0^3} \ {1\over N^2}\ ,\ 
$$
where
$$
u_0={1\over 3r_0} \ , \ \ \ \lambda=g^2_{\rm YM}N\ .
$$
The coupling is of order $1/N$.
The glueball spectrum is obtained by solving the equations of motion for 
the string fields in this background. Note that the metric is independent of $N$,
which shows that to leading order in $1/N$ the glueball spectrum will be independent
 of $N$, as expected in large $N$ theories.
It can be shown that the model exhibit confinement in the form of an area law for Wilson loops
\cite{witt}.
An explicit Wilson loop calculation --based on the Nambu-Goto action for the string dynamics--
shows that the potential between quarks and antiquarks is linearly increasing at large distances
(see \cite{sonn} and references therein). Some other interesting features of this model were explored e.g. in
\cite{gross,csaki,KKglue,HO}.
Below we will study the glueball spectrum.

We have obtained this model as Kaluza-Klein reduction of $4+1$ dimensional
super Yang-Mills theory, which as a quantum field theory is non-renormalizable.
The 3+1 dimensional description applies in a regime where the masses $M_{KK}$ of  Kaluza-Klein
particles with non-zero momentum components along $S^1$ are much larger than the glueball masses. These are typically of order of the string tension
$M_{\rm glue}^2\sim \sigma $, which in the present case  is 
$\sigma = {4\over 3}\lambda u_0^2 $.
On the other hand, Kaluza-Klein particles have masses 
$$
M_{KK}={1 \over r_0} \sim u_0\ .
$$ 
Demanding
this to be much smaller than $M_{\rm glue}$ implies that $\lambda \to 0$.
However, 
this is a regime where the supergravity approximation does not apply. To keep $e^\phi $
fixed in this limit requires $u_0\to 0$, which lead to a  metric which is singular at $u=0$
(in fact, the extremal D4 brane metric).
A generalization of this model based on the rotating D4 brane was proposed in \cite{russo},
and investigated in detail in \cite{csa}, \cite{mina}, \cite{sfet}.
This allows to decouple the Kaluza-Klein states associated with the $S^1$ direction,
but, as discussed below, there still remain extra unwanted Kaluza-Klein states.
Other approaches to QCD using supergravity  can be found in \cite{GPPZ,mathur}.

%%%%%%%%%%%%%%%%%%%%%%%%%%%%%%%
\subsection{Glueball spectrum}
%%%%%%%%%%%%%%%%%%%%%%%%%%%%%%%

Glueball states are conventionally represented  by $J^{PC}$, where $J$ is the spin, and
$P,C$ denote parity and charge conjugation quantum numbers.
Consider the scalar glueball $0^{++}$. The lowest dimension operator with 
$0^{++}$ quantum numbers is ${\cal O}=\tr F_{\mu\nu}^2 $.
The supergravity mode that couples to this operator is the dilaton field fluctuation $\tilde \phi $.
This follows from the D-brane action, which is of the form $I\sim \int d^4 x e^{-\phi}\tr F_{\mu\nu}^2 +...$.
Glueball masses are obtained e.g. by looking for particle poles in correlators
$\langle {\cal O} {\cal O} \rangle $. From the prescription
\be
\langle \exp\bigg[-\int d^4 x \tilde \phi_0(x) {\cal O}\bigg]
 \rangle =e^{-I_{\rm SG}(\tilde \phi _0 )}
\ee
$$
I_{\rm SG}(\tilde \phi _0 )=\int d^{10}x \sqrt{g} e^{-2\phi } \tilde \phi \nabla^2 \tilde\phi +...
$$
it follows that masses will be determined by the eigenvalues of the equation
\be
\p_\mu \big[ \sqrt{g} e^{-2\phi } g^{\mu\nu}\p_\nu \tilde \phi \big]=0\ .
\ee
Solutions are of the form
\be
\tilde\phi =\varphi (u) e^{ik.x} Y(\Omega_4) \ ,
\ee
with $k_\mu $ being the momentum in ${\bf R}^4$ and $Y(\Omega_4)$ the spherical harmonic
of the four-sphere $\Omega_4 $.
The boundary conditions are as follows:

\smallskip
\noindent 
i) at the lower endpoint $u=u_0$ we must demand $\p_u \varphi =0$.

\smallskip
\noindent 
ii) at $u=\infty $ there are two independent solutions, $\varphi \sim {\rm const.}$
and $\varphi \sim u^{-6}$. To have a normalizable solution one must require
$\varphi \sim u^{-6}$. 

\smallskip
\noindent 
As a result, the spectrum is discrete.

Consider for example $SO(5)$ singlets, $\tilde \phi =\varphi (u) e^{ik.x}$.
Using the metric  and dilaton given in eq.~(\ref{metro})  we obtain
\be
{1\over u^3} \p_u\big[ u (u^6-u_0^6)\p_u\varphi \big]= -M^2 \varphi (u)\ ,\ \ M^2=-k^2
\ee
This equation was solved numerically in  \cite{csaki}.
In this way  one finds 
supergravity predictions for masses of  the $0^{++}$ glueball and
its resonances.

Non-singlet $SO(5)$ states have no analogue in QCD since there is no $SO(5)$
global symmetry in pure QCD. Thus those states should decouple in the limit
$\lambda\to 0$ \cite{KKglue}.
 As mentioned above, also Kaluza-Klein states with momentum in the fifth dimension
$\theta $ should decouple in order for the theory to be $3+1$  (rather than $4+1$)
dimensional.
However, the QCD model (\ref{metro}) 
has a single scale in the geometry $u_0$ and, to the leading supergravity approximation,
all masses
are of the same order.
The parameter $\lambda $ enters in next-to-leading order supergravity calculations, giving
corrections to masses which are suppressed by powers  of $1/\lambda $. The supergravity approximation does not apply in the limit $\lambda\to 0$, where the decoupling of extra states should occur.
Solving the full string theory on this background for small $\lambda $ should provide a good 
description of large $N$ pure QCD with no extra unwanted particles, but this is a difficult problem.

It is therefore of interest to look for more general supergravity models of QCD which
can be more effective and useful in computing glueball masses.
The idea is to look for geometries with the same asymptotics and same D-brane charges.
We can think of them as adding ``irrelevant" deformations to the Yang-Mills lagrangian, so that
the theory is in the same universality class, but Kaluza-Klein states are heavy and decouple.
No hair theorems imply that the most general model of this kind (i.e. based on a
{\it regular} geometry with only D4 brane charge) is obtained from a rotating D4 brane parametrized by charge, mass, and two angular momenta.
The corresponding  models were investigated in \cite{russo}, \cite{csa}, \cite{mina}, \cite{sfet},
and the spectra of $0^{++}$ and $0^{-+}$ glueballs and their resonances were determined in the full two-parameter space \cite{sfet}.
The  two extra parameters of these models originate from the angular momenta of the D4 brane. These models include Witten model (\ref{metro}) as a special case.
For large values of these parameters Kaluza-Klein states of $S^1$ are heavy and decouple.
Comparing to the results obtained in lattice QCD, one finds a very interesting agreement \cite{csa}.
Using as input the lattice value of the $0^{++}$ glueball mass, the mass of the first resonance
$0^{++*}$ is 2.55, to be compared with the lattice calculation of $2.8$.
For the $0^{-+}$ state and the resonance $0^{*-+}$, one finds masses equal to 2.56
and  3.49, respectively, which are very close to the lattice values $2.59\pm 0.13$
and $3.64\pm 0.18$ (the simplest model (\ref{metro}) gives 2.00 and 2.98).
The lattice values are for $N=3$, and 
supergravity results are expected to receive corrections of order $1/N^2$.

Another interesting physical quantity that can be computed in this model is the gluon condensate.
{}From the relation $Z(T)=e^{-{F/T}}$, where $F$ is the free energy of the supergravity background, one has \cite{HO}
\be
\langle {1\over 4 g^2_{\rm YM}} \tr F^2_{\mu\nu}(0)\rangle =- {F\over VT}\ .
\ee
The free energy of the background can be computed by the usual
formulas of black hole thermodynamics. In this way one finds \cite{csa}
\be
\langle {1\over 4 g^2_{\rm YM}} \tr F^2_{\mu\nu}(0)\rangle ={1\over 12\pi }\ {N^2\over\lambda }
\ \sigma ^2\ ,
\ee
where $\sigma $ is the string tension. This formula is universal in the sense that it does not depend
on the angular momentum parameters.

One can also compute the topological susceptibility $\chi_t $, which is a measure of
the fluctuations of the topological charge of the vacuum,
\be
\chi_t={1\over (16\pi ^2)^2 } \int d^4x \langle \tr F\tilde F(x)\ \tr F\tilde F(0)\rangle \ .
\ee
The Witten-Veneziano formula \cite{weta,veta} relates the topological suceptibility
of $SU(N)$ Yang-Mills theory without matter fields to the mass of the $\eta' $ boson in $SU(N)$ Yang-Mills theory with $N_f$ quarks,
$M^2_{\eta'} ={4 N_f\over f_\pi^2}\chi_t $. The supergravity calculation gives
an expression of the form $\chi_t \sim \lambda \sigma ^2 $ \cite{HO,csa}.

%%%%%%%%%%%%%%%%%%%%%%%%%%%%%%%%%%%%%
\section{Conclusion}
%%%%%%%%%%%%%%%%%%%%%%%%%%%%%%%%%%%%%

Let us summarize the salient features of the specific
string models of large $N$ gauge theories considered here.
These gauge theories can be described by  string theories, where 
strings fluctuate in higher dimensions.
In the Yang-Mills field theory description, the existence of such extra dimensions is reflected in infinite towers of operators associated with radial modes and Kaluza-Klein states of the supergravity description.

In general, for $N\gg1 $, we expect that any gauge theory should have a string theory description,
though it may not have a classical supergravity description. The supergravity approximation
can be justified provided  curvatures and dilaton coupling $e^\phi $ are small everywhere.
Whenever the gravity solution  contains the $AdS_{p+2}$ space on some slices,
the dual field theory of the boundary will inherit the $SO(2,p+1)$ symmetry group and,
consequently, it will be a conformal field theory.

Many interesting quantities can be calculated in the regime $\lambda \gg 1$, where
the supergravity approximation applies.
These include correlation functions, spectrum of operators or states, Wilson loops
and thermal properties.
The specific QCD model obtained by reduction of $4+1$ dimensional SYM theory
reproduces many qualitative aspects of QCD, but in the supergravity approximation
the model contains extra light states.

Some open questions include a formulation of holography in Minkowski space,
using the correspondence to address the information problem of black holes,
and calculations with controlled approximations in QCD.

\end{document}